\documentclass[twocolumn]{aastex631}

\shorttitle{Circumbinary Rings}
\shortauthors{Betancourt et al.}
\usepackage{amsmath}
\usepackage{units}
\usepackage{mathrsfs}
\usepackage{xcolor}

\begin{document}

\title{Eccentric Disks from Gaseous Rings around Equal-Mass, Circular Binaries}

\author[0009-0008-4616-1527]{Leonardo Betancourt}
\affiliation{Center for Cosmology and Particle Physics, Physics Department, New York University, New York, NY 10003, USA}
\affiliation{Department of Astronomy, California Institute of Technology, 1200 East California Boulevard, Pasadena, CA 91125, USA}

\author[0000-0002-0106-9013]{Andrew MacFadyen}
\affiliation{Center for Cosmology and Particle Physics, Physics Department, New York University, New York, NY 10003, USA}

\author[0000-0002-1895-6516]{Jonathan Zrake}
\affiliation{Department of Physics and Astronomy, Clemson University, Clemson, SC 29634, USA}

%% Note that the \and command from previous versions of AASTeX is now
%% depreciated in this version as it is no longer necessary. AASTeX 
%% automatically takes care of all commas and "and"s between authors names.

%% AASTeX 6.31 has the new \collaboration and \nocollaboration commands to
%% provide the collaboration status of a group of authors. These commands 
%% can be used either before or after the list of corresponding authors. The
%% argument for \collaboration is the collaboration identifier. Authors are
%% encouraged to surround collaboration identifiers with ()s. The 
%% \nocollaboration command takes no argument and exists to indicate that
%% the nearby authors are not part of surrounding collaborations.

%% Mark off the abstract in the ``abstract'' environment. 
\begin{abstract}
We perform high-resolution, grid-based hydrodynamics simulations of gaseous rings viscously spreading into disks around equal-mass, circular binaries. We find that all systems suppress accretion onto the binary when the gas is relatively cold. Circumbinary rings (CBRs) display weak variability above the binary orbital frequency $\Omega_b$ and a dominant spectral peak at $\sim0.1\Omega_b$ (half the fiducial lump frequency of $\sim0.2\Omega_b$). The evolution of CBR eccentricity depends strongly on both the initial ring radius and gas temperature, with smaller, colder rings exhibiting higher eccentricity up to $e \simeq 0.7$. Cold, compact rings develop nearly radius-independent eccentricity profiles, maintaining large $e$ out to several times the initial gas semimajor axis. We find that eccentricity growth favors a stream impact mechanism, in which gas torqued by the binary at pericenter passage exerts a perturbative force on the cavity wall. We consider inefficiently-accreting, intermediate-mass ($\sim10^4 M_\odot$) black hole binaries as sources of quasi-periodic eruptions when rejected streams shock the cavity wall and radiate in the UV or soft X-ray. We discuss the implications of eccentric disks evolved from CBRs for quasar light curves and asymmetric, time-variable double-peaked line emission from disks in galactic nuclei. If binaries drive asymmetry in accretion disk line profiles, our study suggests that the progenitor CBR must have been very compact.

\end{abstract}

%% Keywords should appear after the \end{abstract} command. 
%% The AAS Journals now uses Unified Astronomy Thesaurus concepts:
%% https://astrothesaurus.org
%% You will be asked to selected these concepts during the submission process
%% but this old "keyword" functionality is maintained in case authors want
%% to include these concepts in their preprints.

%% From the front matter, we move on to the body of the paper.
%% Sections are demarcated by \section and \subsection, respectively.
%% Observe the use of the LaTeX \label
%% command after the \subsection to give a symbolic KEY to the
%% subsection for cross-referencing in a \ref command.
%% You can use LaTeX's \ref and \label commands to keep track of
%% cross-references to sections, equations, tables, and figures.
%% That way, if you change the order of any elements, LaTeX will
%% automatically renumber them.
%%
%% We recommend that authors also use the natbib \citep
%% and \citet commands to identify citations.  The citations are
%% tied to the reference list via symbolic KEYs. The KEY corresponds
%% to the KEY in the \bibitem in the reference list below. 

\section{Introduction} \label{sec:intro}
Binary star and black hole systems are expected, on observational and theoretical grounds, to be ubiquitous. Spectroscopic surveys (\citealt{Duquennoy:1991}) as well as astrometric measurements  (\citealt{Raghavan:2010}; \citealt{El-Badry:2021}) suggest that a large number of stars, and the majority of solar-like stars, exist in pairs or higher-order multiples (see \citealt{Duchene:2013} for review). During hierarchical structure formation, galaxy mergers are expected to frequently produce super-massive black hole binaries (SMBHBs) at their centers. Their existence is supported by both analytic (\citealt{Begel:Blan:Rees:1980}; \citealt{Volonteri+2003}; \citealt{MerrittMilos:2005:LRR}; \citealt{Dosopoulou:2017}), and numerical (\citealt{Dotti:2007}; \citealt{Qian:2024}) studies which demonstrate that dynamical friction against the background mass in the galaxy (gas+stars+DM) causes the black holes to migrate to the galactic center, shrinking their orbital separation down to the parsec scale.

Recent hydrodynamics simulations have shown that the time-dependent potential of the binary may produce alternative circumbinary gas morphologies with fundamental differences to the fiducial extended disk model (\citealt{Farris:2014}, \citealt{Munoz:2019}, \citealt{Duffell:2024}). For instance, \cite{Goicovic-I:2016} showed that low-angular momentum clumps of gas from the interstellar medium can form nearly-circular rings of gas around a binary as a result of their near-radial infall. These rings have radii on the order of the binary separation and have been shown to form with random angular momentum vectors with respect to that of the binary, including both prograde and retrograde configurations. Similarly, numerical simulations of tidal disruption events (TDEs) suggest the formation of rings around SMBHBs when a star falls below either BH's tidal radius (\citealt{Coughlin:2016}; \citealt{Coughlin+2017}; \citealt{Ryu:2022}; \citealt{Yu:2025}).

Circumbinary rings are not limited to massive black holes binaries, but also form around binary star systems.  Observations of young stellar objects in the Class II stage of star formation indicate mass accretion rates from protostellar disks in the range of $10^{-12}$--$10^{-7}\ M_\odot\ \mathrm{yr^{-1}}$ (see Fig. 4 in \citealt{Manara:2023}). Given disk lifetimes of $\lesssim 10^7$ yr, typical accretion rates appear insufficient to grow stellar masses. This motivates modes of episodic accretion with short-lived periods of enhanced inflow. Hydrodynamics simulations of young stellar clusters show that turbulent gas can generate misaligned accretion flows whose angular momenta cancel, enabling bursty accretion (\citealt{Bate:2003:StarCluster}; \citealt{Grudic:2022}). In such environments, collisions between stars and turbulent gas facilitates the formation of binary systems surrounded by gaseous rings (\citealt{Borchert:2022}). Ring accretion may have also been observed in ALMA studies of protoplanetary structures. For example, \cite{Guerra-Alvarado:2025} found that two-thirds of stellar systems in Lupus are surrounded by compact dust structures, some in ring-like shapes with radii as small as 0.6 au.

Recent hydrodynamic studies have discovered a regime of suppressed accretion from cold circumbinary disks (CBDs) with characteristic scale height $h/r \sim \mathcal{M}^{-1} \lesssim 0.025$ where $\mathcal{M}$ is the orbital Mach number (\citealt{RagusaLodato:2016}; \citealt{Dittmann:2022:MachStudy}; \citealt{Tiede:2025}). These models, however, assume an `infinite' disk that extends far out from the binary, and it remains to be studied whether suppressed accretion is sensitive to finite-disk effects. \cite{Munoz:2020} studied accretion from relatively warm finite tori and infinite disks with $h/r=0.1$, finding that both disk configurations exert similar torques on the binary. Here, we extend this analysis to determine whether the suppressed accretion regime is robust with respect to disk extent.

We perform high-resolution hydrodynamics simulations of a viscous ring of gas as it relaxes into an accretion disk around a binary over multiple viscous timescales $t_{\rm visc}\sim R_0^2/\nu$, where $R_0$ is the initial ring radius and $\nu$ is the kinematic viscosity. Since all of our simulations are scale-free, our findings are applicable to both stellar and black hole binaries.

This work is organized as follows. In Section \ref{Numerical_Setup}, we detail the initial conditions and numerical methods employed to run our suite of simulations. In Section \ref{Accretion_Periodicity}, we compare accretion signatures between circumbinary gas models. In Section \ref{Gas_Eccentricity}, we report on the growth and saturation of the disk eccentricity and its dependence on our simulation parameters $R_0$, $\mathcal{M}$ and disk extent. In Section \ref{Discussion}, we discuss the observational implications of our results in the context of EM-dark binaries in quasars, quasi-periodic eruptions from galactic centers and double-peaked line emission from accretion disks. In Section \ref{Summary}, we summarize our findings.

\section{Numerical Setup} \label{Numerical_Setup}
{
We solve the 2D, vertically-integrated viscous hydrodynamics equations in Cartesian coordinates. We simulate a gaseous ring surrounding an equal-mass ($q_b=M_2/M_1=1$), circular binary ($e_b=0$) with total mass $M = M_1 + M_2$ and semimajor axis $a$, where $M_1 = M_2$ are the binary component masses. The ring is locally isothermal and geometrically thin with orbital Mach number $\mathcal{M}\sim (h/r)^{-1} \gg 1$.

We prescribe a temperature profile in the disk with a sound speed given by $c_s^2 = -\Phi_b / \mathcal{M}^2$. The binary's potential is implemented with a softened Newtonian potential,
\begin{align}
    \Phi_b = \Phi_1 + \Phi_2,\ \Phi_j = \frac{GM_j}{\sqrt{r_{ij}^2 + \epsilon^2}},
\end{align}
with a gravitational softening length $\epsilon = 0.05a$. Here, $r_{ij}$ is the distance between a computational zone $i$ and binary component $j$.

Viscous stresses are implemented via a constant-$\nu$ viscosity prescription $\nu = \bar{\nu}a^2 \Omega_b$ with $\bar{\nu}=10^{-3}$, where $\Omega_b$ is the binary orbital frequency. Accretion onto each binary component is modeled via torque-free sink terms which remove mass and momentum from the computational domain (see \citealt{Dempsey-TFsinks:2020}). This is the standard circumbinary setup described in \cite{Duffell:2024} but with a modified initial surface density. The gas is initialized as a ring of radius $R_0$ and width $\sigma=0.1R_0$, implemented with a Gaussian radial profile,
\begin{align}
    \Sigma(r) = \Sigma_0 e^{-(r - R_0)^2 / 2\sigma^2},
\end{align}
where $\Sigma_0$ is set such that the initial mass of the ring $M_{\rm ring}= 1$. Note that this is purely for convenience in post-processing, since we implicitly assume $M_{\rm ring} \ll M$ by ignoring self-gravity. We impose a surface density floor of $\Sigma / \Sigma_0 \geq 10^{-10}$ throughout the simulation. Tests with several alternative floor values show that this choice ensures numerical stability without affecting the hydrodynamic evolution. The ring is initialized on a circular, Keplerian orbit with angular frequency $\Omega_0^2 = GM/r^3$.}

We run a total of 25 simulations comparing ring and disk initial conditions. We perform 20 ring simulations exploring radii $R_0 = \{1a, 2a, 3a, 4a\}$ and orbital Mach numbers $\mathcal{M}= \{10, 20, 30, 40, 60\}$. For direct comparison between ring and infinite disk, we run an additional suite of 5 infinite disk simulations at constant-$\nu$ viscosity and equivalent Mach numbers.

The computational domain is square with side lengths $40a$ (ring) and $24a$ (infinite disk) and uses Cartesian coordinates. The resolution is uniform with $\Delta x = 0.01a$. The simulations are performed using \texttt{Meena}, a second-order, GPU-accelerated Godunov hydrodynamics code written in \texttt{JAX} (\citealt{jax2018github}) by the lead author.

\section{Accretion Dynamics}\label{Accretion_Periodicity}
\subsection{Suppressed Accretion}\label{suppression}
\begin{figure*}
    \centering
    \includegraphics[width=\textwidth]{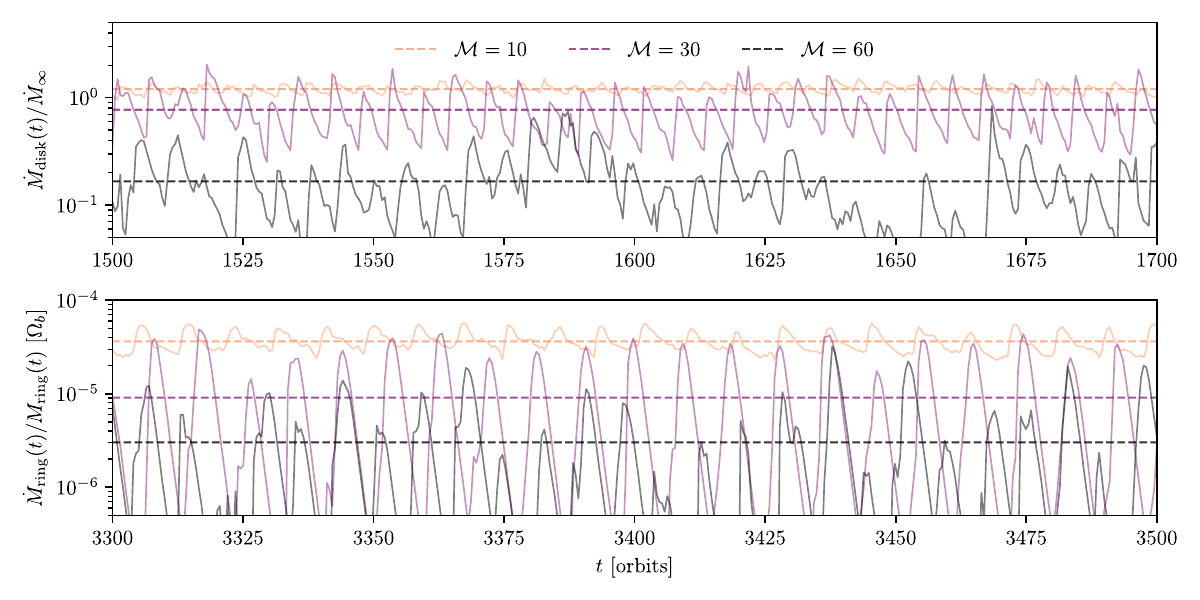}
    
    \caption{(Top) Accretion rate timeseries of the infinite disk in quasi-steady-state, normalized by the large-scale inflow rate, $\dot{M}_\infty=3\pi\Sigma_0\bar{\nu}a^2\Omega_b$. (Bottom) Accretion rate timeseries of the $R_0=4a$ ring well after $t_{\rm visc}$. In both plots, the dashed lines represent the time-averaged normalized accretion rates.}
    \label{fig:m_dot_panels}
\end{figure*}

\begin{figure}
    \centering
    \includegraphics[width=\linewidth]{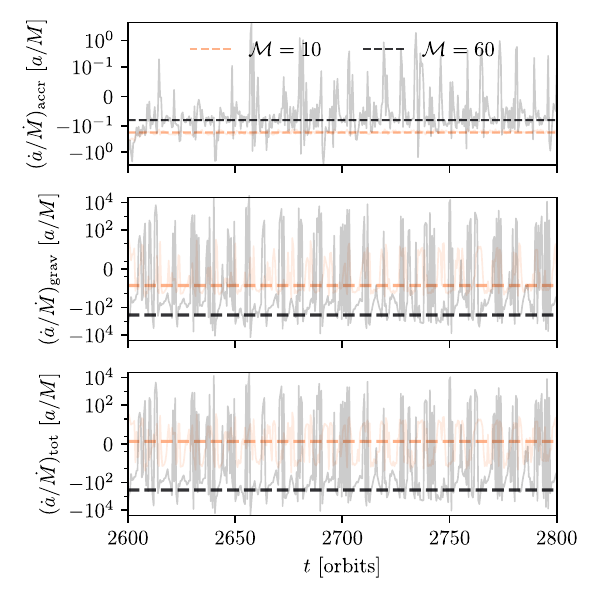}
    
    \caption{Accretion (top), gravity (middle), and total (bottom) contributions to the time-derivative of the binary semimajor axis $a$.
    }
    \label{fig:a_dot_panels}
\end{figure}
In the initial transient period, all rings studied become tidally disrupted by the binary potential. Gas in the outer ring spreads radially outward, while gas in the inner edge is pulled into streams, leading to capture by either binary component or flinging of the stream back out to the edge of the cavity (the central region around the binary containing very low density of fluid; \citealt{MacFadyen:2008}). While the ring exhibits complex angular momentum redistribution during this transient period, we focus here on the feeding dynamics of a relaxed disk evolved from a ring initial condition. 

Figure \ref{fig:m_dot_panels} compares quasi-steady-state accretion rates of the `infinite' disk and the $R_0=4a$ ring following the initial transient period. In the disk case, $\dot{M}$ is normalized by the large-scale inflow rate, $\dot{M}_\infty=3\pi\Sigma_0\nu a^2\Omega_b$ (see \citealt{Tiede:2025}). Rings have finite extent and therefore don't have an equivalent large-scale inflow rate. After a viscous time, we observe the relaxed ring to accrete at a rate that decreases as it is secularly depleted (\citealt{Munoz:2020}). Due to this depletion on the viscous timescale, we normalize the reported accretion rates by the time-dependent mass of the ring. Therefore, $\dot{M}_{\rm ring}(t) / M_{\rm ring}(t)$ gives a mass- and time-independent measure of accretion rate that we can compare at different $\mathcal{M}$. We measure accretion suppression in both infinite disks and rings as the ratio of the accretion rate at $\mathcal{M}=10$ and successively higher $\mathcal{M}$ for a given initial condition. In the top panel of Figure \ref{fig:m_dot_panels}, we recover the accretion suppression observed previously in the limit of a thin infinite disk, where the  time-averaged accretion rates decrease monotonically with increasing $\mathcal{M}$. The $\mathcal{M}=60$ disk accretes at a rate $\sim 10\%$ of the $\mathcal{M}=10$ disk. The bottom panel shows a very similar suppression factor in the relaxed ring of relatively cold ($\mathcal{M}=60$) gas. Both infinite disks and finite rings exhibit equal accretion suppression for gas at higher $\mathcal{M}$. In Figure \ref{fig:m_dot_panels} we report normalized accretion rates for the $R_0=4a$ ring, but our simulations suggest that the level of suppression is independent of initial ring radius. This suggests that accretion dynamics are insensitive to large-scale gas structure and are instead driven largely by interaction between the binary and the cavity inner edge. In other words, the efficiency of stream capture onto the binary component decreases dramatically in colder gas, regardless of initial condition.

Figure \ref{fig:a_dot_panels} shows the individual contributions to the evolution of the binary separation $\dot{a}$, as well as their combined effect. For an equal-mass binary, this rate is given by
\begin{align}
    \frac{\dot{a}}{\dot{M}} = 2\left(\frac{\ell_0}{\ell} - \frac{3}{2}\right) \frac{a}{M},
\end{align}
where $\ell_0\equiv \dot{L}/\dot{M}$ is the total torque on the binary normalized by the accretion rate and $\ell\equiv L/M$ is the specific angular momentum of the binary. Figure \ref{fig:a_dot_panels} shows that the evolution of the binary separation is dominated by gravitational torques. In colder disks, streams of gas are more likely to miss the binary, effectively stealing specific angular momentum on the order of $a^2\Omega_b$ with each binary encounter (\citealt{Tiede:2025}). The result is that cold disks and rings alike push the binary closer together, with increasingly colder gas driving faster inspiral.

In comparison with previous studies by \cite{Tiede:2025}, we find that the level of suppression is sensitive to the viscosity presciption in the disk. \cite{Tiede:2025} showed $\alpha$-disks with $\bar{\nu}=10^{-4}$ to exhibit a $\sim 1\%$ suppression between $\mathcal{M}=10$ and $\mathcal{M}=60$, while our findings suggest a $\sim10\%$ suppression factor. However, since accretion disks are likely much thinner, in the range $h/r\sim \mathcal{M}^{-1}\sim 10^{-2}-10^{-3}$ (\citealt{SS73}), the astrophysical consequences are robust. Cold CBDs may produce reduced accretion-driven electromagnetic luminosity, regardless of their radial extent. 

\subsection{Accretion Periodicity}
\begin{figure*}
    \centering
    \includegraphics[width=\textwidth]{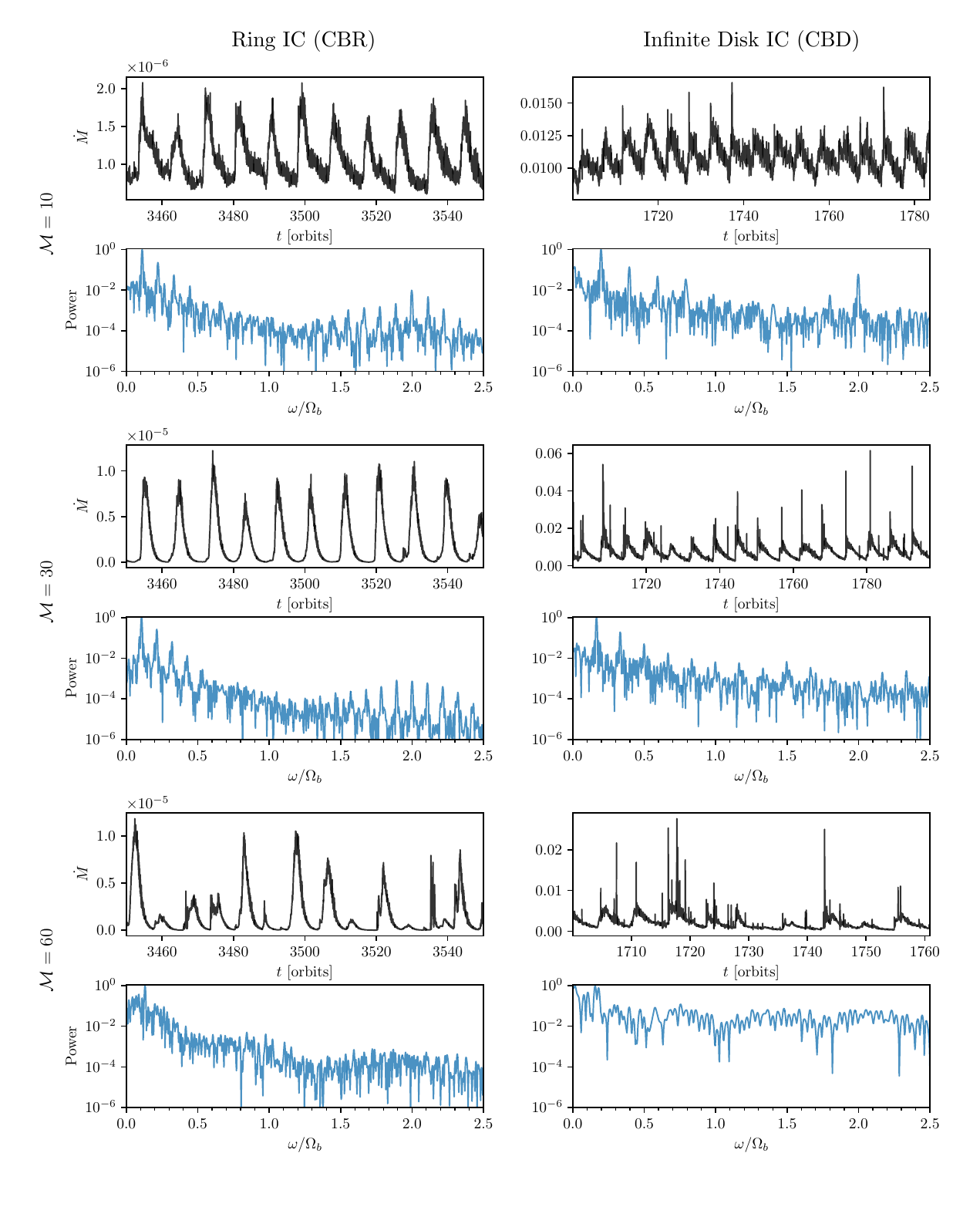}
    
    \caption{100-orbit windows of the accretion rate (top) and the corresponding Lomb-Scargle periodogram (bottom) computed over 100 orbits in quasi-steady-state. Left and right panels compare ring and infinite disk initial conditions (ICs) respectively.}
    \label{fig:m_dot_lombscargle}
\end{figure*}
As the tidal potential drives an eccentric instability in the cavity, periodic `lump' accretion emerges. Rejected streams pile up coherently on the opposite edge of the cavity wall, producing an over-density that delivers mass to the binary at a certain frequency (\citealt{Lubow:1991:Model}; \citealt{MacFadyen:2008}; \citealt{Shi:2012}). Figure \ref{fig:m_dot_lombscargle} shows 25-orbit timeseries of the accretion rate, which is modulated by this periodic signal. At $\mathcal{M}=10$, the circumbinary ring produces the familiar sawtooth accretion pattern observed previously in simulations of the infinite disk. For higher $\mathcal{M}$, however, we observe accretion signatures that differ depending on initial conditions. In the infinite disk, the sawtooth pattern is robust for varying $\mathcal{M}$. In the ring, accretion flares become significantly more symmetric and are closer to triangular wave patterns as $\mathcal{M}$ increases. Figure \ref{fig:m_dot_lombscargle} also demonstrates the discrepancy in accretion flare amplitudes associated with suppressed `lump' delivery for $\mathcal{M}=60$. Inefficiency of stream capture at $\mathcal{M}=60$ makes the overdensity more likely to miss the binary if it hasn't yet accumulated enough gas from rejected streams.

In the fiducial CBD, `lump' accretion episodes occur at a frequency of $\sim0.2 \Omega_b$ (every $\sim 5$ binary orbits; \citealt{MacFadyen:2008}). We find a weak dependence of this dominant frequency on $\mathcal{M}$, with higher $\mathcal{M}$ disks exhibiting slightly lower `lump' frequencies. In the circumbinary ring, the dominant frequency is significantly lower, at $\sim 0.1\Omega_b$. Unlike the infinite disk, this dominant frequency appears to be largely insensitive to $\mathcal{M}$. 

Beyond the dominant frequency, the character of the periodograms differ significantly between initial conditions. In the infinite disk, subdominant frequencies at $\sim 0.4\Omega_b$ and $\sim 2\Omega_b$ become lost to noise for higher $\mathcal{M}$, with $\mathcal{M}=60$ exhibiting approximately constant power for all frequencies above the dominant peak. In the ring, these subdominant frequencies appear much weaker, and their strength diminishes for higher $\mathcal{M}$. At $\mathcal{M}\gtrsim 40$, frequency noise above the dominant frequency has two orders of magnitude less power than the infinite disk at equivalent $\mathcal{M}$. Thus, CBRs exhibit significantly cleaner periodicity with a single dominant mode at $\sim 0.1 \Omega_b$ and accretion variability with a quasi-triangular shape.

\section{Gas Eccentricity}\label{Gas_Eccentricity}
\subsection{Time Evolution}
\begin{figure*}
    \centering
    \includegraphics[width=\textwidth]{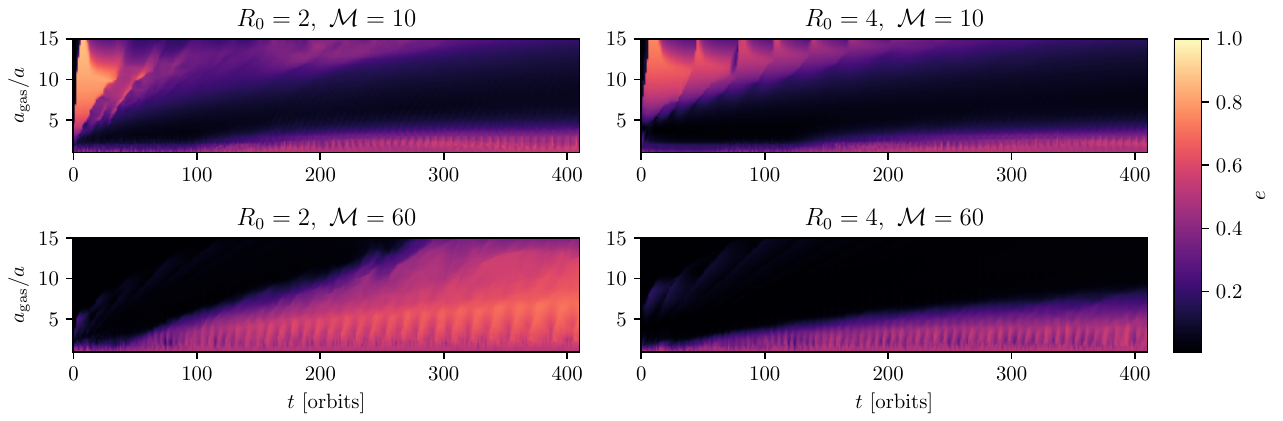}
    
    \caption{Spacetime plots showing the time evolution of gas eccentricity $e$ averaged and binned by gas semimajor axis $a_{\rm gas}$. During the initial tidal disruption of the ring, the binary unbinds gas that is flung out to large semimajor axis (top left part of each plot), eventually flowing out of the domain through the outer boundary. Gas that remains bound in a ring grows in $e$ out to a semimajor axis that depends on $R_0$ and $\mathcal{M}$.}
    \label{fig:e_spacetime}
\end{figure*}

We compute the gas eccentricity in a given computational zone as the magnitude of the Laplace-Runge-Lenz vector (e.g. \citealt{Miranda:2017}; \citealt{MunozLithwick:2020}; \citealt{Siwek:2022}),
\begin{align}
    \textbf{e}_i = \frac{\textbf{v}_i\times (\textbf{r}_i \times \textbf{v}_i)}{GM} - \frac{\textbf{r}_i}{|\textbf{r}_i|}. \label{e_lrl}
\end{align}
Figure \ref{fig:e_spacetime} illustrates the time evolution of the magnitude of $\textbf{e}$ binned by gas semimajor axis $a_{\rm gas}$. At early times, gas on the ring inner edge is perturbed by the binary potential. This gas undergoes strong gravitational interactions with the binary and a portion of this gas becomes unbound, traveling out to large radius and out of the computational domain, as seen by the overdensity of gas at large $e$ and $a_{\rm gas}/a$ early in the simulation.

Across all initial conditions, the binary potential drives an instability in the orbiting gas and leads to the growth of the eccentric cavity (\citealt{MacFadyen:2008}). Rings at $\mathcal{M}=10$ remain largely circular beyond $a_{\rm gas}\simeq 2.5a$, which corresponds to the boundary of the low-density cavity (Figure \ref{fig:e_spacetime}, top panels). For rings at $\mathcal{M}=60$, circumbinary gas grows to greater eccentricity and out to larger semimajor axis (Figure \ref{fig:e_spacetime}, bottom panels). Rings of initially smaller extent $R_0$ at high $\mathcal{M}$ exhibit this behavior most dramatically, with highly eccentric gas orbits occupying the entire space of $a_{\rm gas}$ (Figure \ref{fig:e_spacetime}, bottom left).

\begin{figure}
    \centering
    \includegraphics[width=\linewidth]{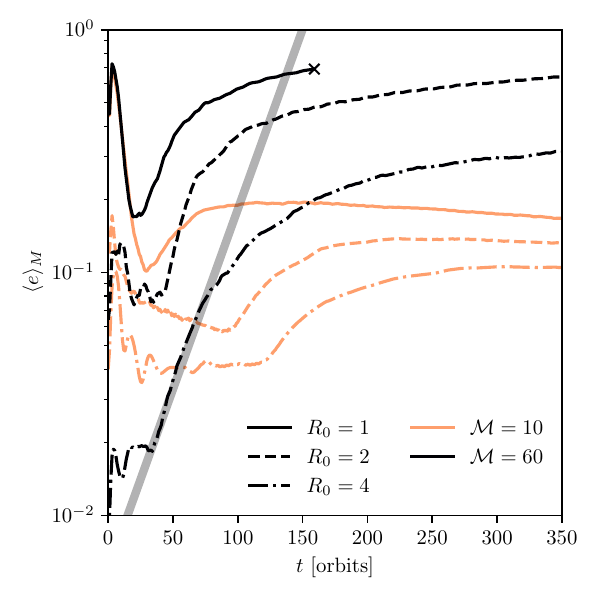}
    
    \caption{Time evolution of the magnitude of the mass-weighted eccentricity vector $\langle e\rangle _M$, defined in Eq. \eqref{e_M}.  The thick gray line represents exponential growth of $e$ with constant gr owth rate $\gamma_e \simeq 5.5\times 10^{-3}\Omega_b$ computed from stream impacts in Section \ref{stream_impacts}. The `X' marks the point in the simulation when the $\{R_0=1a$, $\mathcal{M}=60\}$ ring hits the outer boundary due to its extreme eccentricity.}
    \label{fig:ecc_vs_time}
\end{figure}

To quantify the bulk eccentricity of the CBR, we define a mass-weighted eccentricity vector as
\begin{align}
    \langle \textbf{e} \rangle_M = \frac{1}{M_{\rm ring}(r>a)} \int_{a}^{20a} \int_0^{2\pi} \textbf{e}(r,\phi) \Sigma dA, \label{e_M}
\end{align}
where $M_{\rm ring}(r > a)$ is the mass of the ring outside the vicinity of the binary. We plot this diagnostic in Figure \ref{fig:ecc_vs_time}. CBRs with $\mathcal{M}=10$ undergo an initial period of eccentricity growth but eventually stall, approaching a bulk eccentricity of $e\sim 0.1$. For the duration of our simulations, CBRs at $\mathcal{M}=60$ are driven to higher eccentricities at a faster rate, and appear to grow unbounded in $e$. Figure \ref{fig:ecc_vs_time} marks the point `X' where the $\{R_0=1a, \mathcal{M}=60\}$ ring grows so eccentric it impacts the outer boundary. Thus, from this simulation we report a lower bound in eccentricity.

We posit that eccentricity in circumbinary gas is governed by two competing mechanisms: (1) pressure gradients in warm disks that damp e, and (2) stream impacts that tend to excite it. In warmer gas (at lower $\mathcal{M}$), more frequent collisions between particles preferentially produces large deviations in azimuthal velocity, causing the gas to spread radially (\citealt{Frank:2002}). This has a circularizing effect. By contrast, stream impacts can exchange angular momentum with the inner disk, typically increasing e depending on the impact geometry and relative angular momentum. In Section \ref{stream_impacts}, we compute the eccentricity growth rate due to perturbative stream impacts and show that it agrees with simulations at early times. These two mechanisms may also be coupled: weaker pressure gradients in colder disks can produce thinner tidal streams that are less efficiently captured by the binary, leading to stronger impacts from returning streams \citep{Tiede:2025}.

\subsection{Radial Distribution}
\begin{figure}
    \centering
    \includegraphics[width=\linewidth]{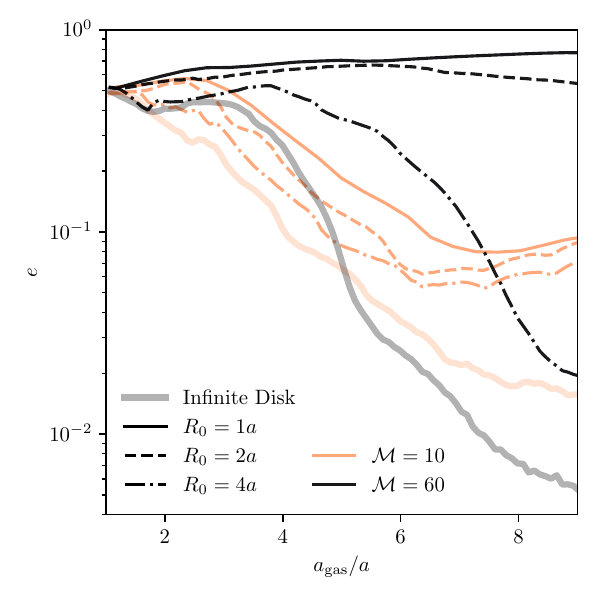}
    
    \caption{Eccentricity $e$ of the circumbinary gas binned by semi-major axis of the gas orbit, $a_{\rm gas}$. The distributions are time-averaged at $t \approx 300$ orbits, when quasi-steady-state is reached (except for $R_0=1a,\ \mathcal{M}=60$, which grows in eccentricity out to the edges of the computational domain after $t\approx 180$ orbits, so the distribution of $e$ is measured at this time).}
    \label{fig:e_profiles}
\end{figure}

 Figure \ref{fig:e_profiles} shows the distribution of $e$ averaged and binned by semi-major axis of the gas orbit, $a_{\rm gas} = -GM/2\epsilon$, where $\epsilon$ is the gas specific orbital energy. Our simulation of the fiducial CBD model at $\mathcal{M}=10$ agrees with the exponential decay of eccentricity with radius reported by \cite{MacFadyen:2008}, following a $e\propto \exp(-a_{\rm gas}/a)$ dependence. All systems converge to an eccentricity of $e\simeq 0.5$ in the vicinity of the binary, where minidisks dominates the local density of fluid \citep{Westernacher:2024}.
 
 At large $a_{\rm gas}$, both the shape and magnitude of the eccentricity distribution are strongly dependent on $R_0$ and $\mathcal{M}$. Rings of smaller initial extent grow to higher eccentricity at all $a_{\rm gas}$, approaching a flat distribution at $\mathcal{M}=60$. Progressively larger rings approach the infinite disk distribution of eccentricity, where $e$ is damped and prevented from propagating out to large radii. 

We note that moderate eccentricities ($e \sim 0.1$–$0.3$) have been found in simulations of infinite CBDs. \citet{MunozLithwick:2020} and \citet{Siwek:2022} report eccentricities up to $e = 0.3$ for gas near the binary. However, this eccentricity decays with radius, typically following a power law or a power law with an exponential cutoff, leaving the outer disk largely circular. In contrast, Figure~\ref{fig:e_profiles} shows that cold, compact CBRs form eccentric rings, with gas remaining on eccentric orbits out to large $a_{\rm gas}$.

\subsection{Stream Impacts}\label{stream_impacts}
A robust dynamical feature of circumbinary disks and rings is the tidal stripping of gas from the cavity wall, creating streams that undergo close encounter with the binary. The efficiency of stream capture onto the binary appears to be very sensitive to $\mathcal{M}$, as discussed in Section \ref{suppression}, and also explored recently by \cite{Tiede:2025}. For higher $\mathcal{M}$,  gas streams are less efficiently captured by the binary. Streams that miss the binary are torqued by the time-dependent potential and rejoin the disk after pericenter passage. These repeated stream impacts induce a perturbative force on the gas orbiting along the cavity edge (see Sec. 4.3.3. in \citealt{Shi:2012}). The perturbation caused by stream impacts is related to the rate of change of the ring eccentricity via the Gauss equations of celestial mechanics (\citealt{Brouwer&Clemence:1961}):
\begin{align}
    \frac{de}{dt} = \frac{\sqrt{1-e^2}}{a_r\Omega(a_r)} \left[ R\sin (f) + S\frac{4\cos(f) + 3e + e\cos (2f)}{2(1 + e\cos(f))}\right]. \label{dedt}
\end{align}
Here, $a_r$ is the semi-major axis of the eccentric ring, $\Omega(a_r)=\sqrt{GM/a_r^3}$ is the Keplerian angular frequency at $r=a_r$, and $f$ is the true anomaly of the stream impact. $R$ and $S$ are the radial and angular components that make up the disturbance force density of the stream,
\begin{align}
    (R, S)=\dot{m_s}\frac{\textbf{v}_s - \textbf{u}_r}{m_r}, \label{force_density}
\end{align}
where $\dot{m_s}$ is the rate of mass injection into the ring, $m_r$ is the mass of the ring, $\textbf{v}_s$ is the velocity of the stream and $\textbf{u}_r$ is the velocity of the ring near impact.

\begin{figure}
    \centering
    \includegraphics[width=\linewidth]{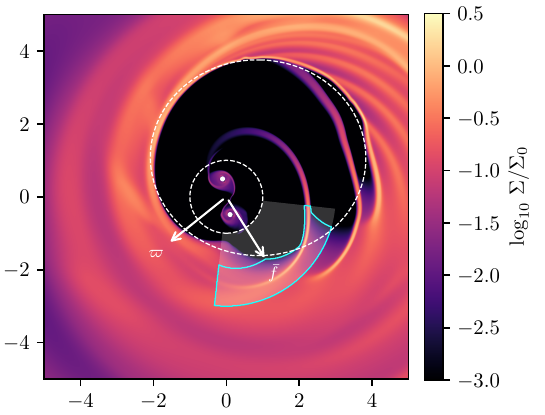}
    
    \caption{Density snapshot of ring $\{R_0=4a,\ \mathcal{M}=60\}$ at $t\approx 300$ orbits, overplotted with various parameters computed during analysis of stream impacts. The dotted line denotes the best-fit ellipse to the shape of the low-density cavity. The fit yields a measurement of the longitude of pericenter $\varpi$, the ring eccentricity $e$ and the semimajor axis $a_r$. The streams are analyzed inside the region bounded by the transparent wedge centered on the mass-averaged true anomaly of stream impacts $f$. The cyan contour denotes the mask used to isolate streams of gas rejected by the binary.}
    \label{fig:stream_snapshot}
\end{figure}

Our simulations suggest that gas streams are created near orbital pericenter and impact along an arc defined by the cavity wall. As the ring becomes more eccentric, the impact is smeared across a larger area of the cavity wall, but to gain an intuitive understanding of this mechanism we can assume the impact is localized to a true anomaly $f\approx \pi / 2$ (see Figure \ref{fig:stream_snapshot}). In this approximation, Eq. \eqref{dedt} simplifies to
\begin{align}
    \frac{de}{dt} = \frac{\sqrt{1-e^2}}{an}\left( R+eS \right) \propto R+eS \label{dedt_exp}
\end{align}
Recalling that $S \propto (\mathbf{v}_s - \mathbf{u}_r)\cdot \hat{\phi}$, we have $S > 0$ ($S < 0$, respectively) when the azimuthal velocity of the stream exceeds (falls below) that of the ring at the impact location. Streams are accelerated by the binary potential and thus carry enhanced specific angular momentum relative to the CBR gas, implying $S > 0$. Similarly, the radial force density satisfies $R \propto (\mathbf{v}_s - \mathbf{u}_r)\cdot \hat{r}$, so that $R > 0$ for streams that are flung outward from the binary. Therefore, streams impacting the cavity wall near $f \approx \pi/2$ generically yield $R + eS > 0$, and thus drive eccentricity growth, $de/dt > 0$. In addition, when $S>0$, Eq. \eqref{dedt} admits solutions of exponential growth in $e$.

To further explore the role of this mechanism in driving eccentricity in CBRs, we numerically compute $de/dt$ following \citet{Shi:2012}, using Eq.~\eqref{dedt} and high-cadence simulation snapshots output at a rate of $60\ \mathrm{orbits^{-1}}$. We focus on the setup $\{R_0 = 4a, \mathcal{M} = 60\}$, which remains well confined within the computational domain and permits analysis of stream impacts in a quasi-steady state. We study the $\mathcal{M} = 60$ case to contrast with the significantly warmer disk studied by \citet{Shi:2012}, which has a characteristic $\mathcal{M} \sim 20$ at $r = a$. This comparison allows us to isolate the effect of disk temperature on eccentricity growth, as mediated by the strength of stream impacts.

Figure~\ref{fig:stream_snapshot} shows a density snapshot of a representative stream impact event, along with orbital parameters measured from the simulation. We fit an ellipse (dotted line) to the low-density cavity to characterize the inner ring, obtaining $a_r/a \approx 4.0$ and $e \approx 0.38$. These values remain approximately constant over the interval $300\Omega_b \leq t \leq 320\Omega_b$ considered here. Streams impact the cavity wall over a broad range of true anomaly from pericenter to apocenter, but the mass-weighted mean is $\bar{f} \approx \pi/2$.

Motivated by this localization, we restrict our analysis to a wedge centered at $f = \pi/2$ with opening angle $\Delta f = \pi/2$. The radial extent of the wedge is $1a \leq r \leq 3a$, capturing the interaction between the stream and the cavity wall as the stream diffuses. This allows us to evaluate Eq.~\eqref{dedt} under the approximation that stream impacts are localized in $f$. Within this wedge, we isolate stream material by applying the following criteria:
\begin{enumerate}
    \item The specific orbital energy satisfies $\epsilon \geq f_\epsilon\,(-GM/a)$, selecting marginally bound material expected for recently ejected streams. We adopt $f_\epsilon = 0.8$.
    \item The surface density satisfies $\Sigma \geq f_\Sigma\,\max(\Sigma)$, isolating dense streams from the background CBR. We use $f_\Sigma = 10^{-3}$.
    \item The radial velocity is positive, $v_r > 0$.
\end{enumerate}

\begin{figure}
    \centering
    \includegraphics[width=\linewidth]{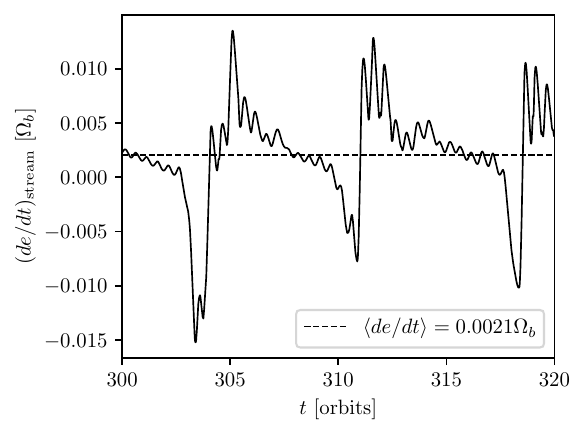}
    
    \caption{Time-derivative of the ring eccentricity $de/dt$ due to stream impacts, computed using snapshots of simulation $R_0=4a,\ \mathcal{M}=60$ at high cadence ($60\ \mathrm{orbit^{-1}}$).}
    \label{fig:dedt_stream}
\end{figure}

Using the resulting stream mask, we integrate the mass injection rate and force density (Eq.~\eqref{force_density}) over the wedge, and compute $de/dt$ via Eq.~\eqref{dedt}. The resulting time series is shown in Figure~\ref{fig:dedt_stream}. Peaks in $de/dt$ coincide with the passage of the lump overdensity through pericenter, which enhances the stream density. We find a time-averaged value $\langle de/dt \rangle \simeq 2.1 \times 10^{-3}\,\Omega_b$, significantly larger than the value reported by \citet{Shi:2012}, $\langle de/dt \rangle \simeq 1.5 \times 10^{-3}\,\Omega_b$. This difference is consistent with colder disks producing less efficient accretion and stronger stream impacts, thereby enhancing eccentricity growth. Assuming exponential growth in $e$ as motivated by Eq. \eqref{dedt_exp}, we find a growth rate
\begin{align*}
    \gamma_e = \frac{\langle de/dt \rangle}{e} \simeq 0.0055\,\Omega_b,
\end{align*}
which we overplot in Figure~\ref{fig:ecc_vs_time} as $e \propto \exp(\gamma_e t)$ (thick gray line). The growth rate measured in quasi-steady state agrees well with the early time evolution of $e$. At later times in the $\mathcal{M}=60$ run, $e$ continues to increase at a reduced rate. This may reflect a decrease in the efficiency of stream-driven forcing as the ring becomes more eccentric: stream impacts are distributed over a broader area of the cavity wall and become less localized, leading to a weaker net perturbative force on the gas.

\subsection{Tidal/Resonant Forcing}
Linear resonance theory predicts that gaseous accretion disks can become eccentric even around circular binaries \citep{Lubow:1991:Model}. The tidal binary potential excites spiral density waves at outer and inner Lindblad resonances (OLR/ILR), and the coupling of wave excitation with the tidal field can drive eccentricity growth. This mechanism has been invoked to explain the superhump phenomenon in cataclysmic variables (\citealt{Goodchild&Ogilvie:2006}), as well as eccentricity in protoplanetary disks with embedded planets (\citealt{Teyssandier:2016}; \citealt{Teyssandier:2017}).

For a circular binary, the dominant azimuthal mode is $m=2$, which excites the 2:1 eccentric Lindblad resonance at $r \simeq 1.6a$ \citep{Lubow:1991:Model}. In the context of CBRs, prior to cavity formation this resonance may facilitate the transition from circular to elliptical gas orbits, particularly in rings with $R_0 \sim 1$–$2a$, where much of the circumbinary material lies near the resonant location. During this initial transient, symmetric streams produce equal and opposite perturbations and therefore can't drive eccentricity as predicted by Eq. \eqref{dedt}.

However, as the cavity expands and becomes elliptical, the gas density near $r \simeq 1.6a$ declines substantially. At higher $\mathcal{M}$, the cavity grows large enough that this resonance lies almost entirely within a low-density region. If eccentricity growth were primarily driven by this tidal resonance, one would therefore expect the growth rate to decrease with increasing $\mathcal{M}$. Instead, we find the opposite trend: higher-$\mathcal{M}$ disks exhibit enhanced eccentricity growth. This discrepancy suggests that tidal resonance alone cannot account for the observed behavior, and instead points to stream impacts as the dominant driver in this regime.

\section{Discussion}\label{Discussion}
\subsection{Binary Periodicity}
The dominant source of EM variability in circumbinary gas systems is believed to originate in the inner disk, where tidal streams deliver gas to each component and accrete through minidisks (\citealt{Cocchiararo_2024}). Detecting such flux variability, assumed to be closely tied to the rate of accretion, may allow us to infer the binary's orbital period and spatial separation. The associated frequency distribution has been shown to vary significantly with the binary orbital eccentricity $e_b$ and mass ratio $q_b$ (\citealt{DOrazio:2024}). Here, we show that this variability is additionally sensitive to large-scale gas distribution. For an equal-mass, circular binary, rings exhibit clean, robust accretion variability at a frequency $\sim 0.1\Omega_b$ (Fig. \ref{fig:m_dot_lombscargle}, left panels); the infinite disk dominant frequency is $\sim 0.2\Omega_b$ (Fig. \ref{fig:m_dot_lombscargle}, right panels). 

If circular binaries are surrounded by infinite disks, their variability may be the more difficult to detect. Lomb-Scargle or FFT analysis performed on short ($\sim$year) intervals of available timeseries data may not produce a high enough signal-to-noise ratio to identify a dominant frequency (\citealt{Cocchiararo_2024}). Figure \ref{Accretion_Periodicity} suggests that periodograms of CBRs may be much cleaner, especially as $\mathcal{M}$ increases. 

Periodic signals from binaries in quasars are likely obscured by the stochastic variability inherent to AGN emission (\citealt{Graham:2017}). Because of this, we caution the use of the Lomb-Scargle periodogram (LSP) to derive periodicity from light curves (LCs) of binary candidates. \cite{Lin:2025} showed that the algorithm struggles with non-sinusoidal time series data representing binary signatures in quasars. In particular, sawtooth-like variability was detected for only a fraction ($1.1\%$) of idealized light curves compared to sinusoidal signals ($28.1\%$). Here we find that the enhanced eccentricity of disks evolved from circumbinary rings contributes to a new mode of accretion variability, resembling a $\sim$triangular wave pattern. The accuracy of the LSP for triangular signatures is yet to be tested.

\subsection{Dark Binary Mergers in Quasars}
We have found that the regime of suppressed accretion onto SMBHBs is robust to varying initial gas distribution in disk and ring initial conditions. As a result, we posit that cold gas may produce significantly weaker EM emission even if radial gas trajectories form a ring of gas at close encounter with the binary. This may limit prospects for detecting an EM-counterpart to SMBHB mergers, as such mergers may occur in the absence of surrounding gas. However, following binary coalescence, viscous inflow of the circumbinary gas is expected to produce a luminous AGN turn-on (\citealt{Milos:Phinney:2005}; \citealt{Tanaka:2010}; \citealt{Shapiro:2010}) that may be observable only a few years after the GW-driven merger (\citealt{TanakaMenou:2010}). Studying the lasting changes in AGN variability in so-called changing-look inspirals (CLIs) may reveal features of circumbinary disk interaction including gas temperature, viscosity and binary mass ratio (\citealt{Zrake:2025}).

We have shown that, as the accretion suppression factor increases, the gas exerts increasingly negative torques on the binary (Figure \ref{fig:a_dot_panels}; in agreement with \citealt{Tiede:2020}). This may imply an efficient channel for the formation of massive black holes from mergers of lower mass (including stellar-mass) black hole binaries enabled by circumbinary disks. Gas-driven mergers, while likely EM-faint, may be a significant contributor to the GW background (\citealt{Agazie:2023}).

\subsection{Quasi-Periodic Eruptions from Intermediate-Mass Black Hole Binaries}
Given the robustness of suppressed accretion in the high-$\mathcal{M}$ regime, we might expect the dominant source of variability to be from the inner CBD instead of accreting minidisks. Our simulations show that, even in this regime, rejected streams can shock heat the cavity wall and radiate in the UV or soft X-ray (\citealt{Westernacher-Schneider:2022}). Recently, a number of galactic nuclei have been detected emitting bursts of soft X-rays in so-called quasi-periodic eruptions (QPEs) with periods ranging from hours to weeks (\citealt{Miniutti:2019}; \citealt{Giustini:2020}; \citealt{Arcodia:2021}; \citealt{Arcodia:2024}; \citealt{Guolo:2024}; \citealt{Nicholl:2024}).  The dominant explanation has been a (stellar) body colliding with the accretion disk on a misaligned orbital plane (e.g. \citealt{Linial:2025}). We propose a variation of the \cite{Nicholl:2024} picture, in which a ring of gas is formed around a binary by tidal disruption of a star or infall of low angular momentum gas, and gas streams rejected by the binary impact the inner cavity wall and radiate quasi-periodically. If QPEs were instead sourced by stream impacts in circumbinary accretion flows, we would expect periodic luminous bursts at twice the binary orbital frequency. 

We can constrain the number of BHBs with QPE-like periods using future LISA event rates $\mathcal{R}_0$, which provide an estimate for the rate of BHB mergers with a given mass. Here we consider intermediate-mass ($M\sim 10^4 M_\odot$) black hole binaries (IMBHBs). From the derivation in Appendix \ref{QPE_calc}, we find that the number of candidate IMBHBs that could be QPE sources is
\begin{align}
    N \approx 4\times 10^7 \left( \frac{\mathcal{R}_0}{1\ \mathrm{yr^{-1}}} \right) \left( \frac{M}{10^4 M_\odot} \right)^{-5/3} \left( \frac{P}{13.3\ \mathrm{days}} \right)^{8/3}, \label{population}
\end{align}
where a period of 13.3 days corresponds to the widest binary with inspiral still dominated by GW emission. For an estimated merger rate of $R_0 \sim 1\ \mathrm{yr^{-1}}$ (\citealt{Fragione:2022}), the scaling given in Eq. \eqref{population} implies a raw population of order $\sim 10^7$ IMBHBs with $M\sim 10^4 M_\odot$ and periods in the QPE range.  Eq. \eqref{population} also estimates a population of $\sim 10^7$ IMBHBs with $M\sim 10^5 M_\odot$ and $P\lesssim 2$ months, which may be more easily detectable by LSST given its lower temporal cadence. 

We note that only a fraction $f_{\rm gas}$ of binaries are surrounded by a gaseous environment, so the number of luminous binaries is better approximated by $f_{\rm gas}N$. However, given the large raw population $N$, the number of detectable binaries remains significant even if $f_{\rm gas}$ is on the order of 1\%.

\subsection{Supermassive Black Hole Binaries in Low-Luminosity Active Galactic Nuclei}
AGN manifest across a wide spectrum of accretion flows, spanning super-Eddington accretion rates to highly inefficient, advection-dominated flows in low-luminosity AGN (LLAGN). Most AGN identified in nearby galaxies have optical signatures of nuclear activity in the form of low-ionization nuclear emission line regions (LINERs) which have been linked to LLAGN accreting far below the Eddington limit (\citealt{Ho:1997}; \citealt{Ho:1999}; \citealt{Terashima:2000}; \citealt{Terashima:2003}; \citealt{Dudik:2009}; \citealt{Gonzalez:2009}). LINERs are produced by hot, collisionally excited species (e.g. O I, O II, S II and N II) with low ionization fractions and thus point to low UV ionizing flux from the central AGN engine (see \citealt{Marquez:2017} for review).

Inefficiently-accreting SMBHBs may have similar observational properties as LLAGN with LINER emission. In these systems, inefficient stream capture starves the accreting minidisks and thus suppresses their UV/optical emission. The lack of luminous minidisks may therefore lead to weak or absent broad line regions and a spectrum dominated by low-ionization transitions. Thus SMBHBs may resemble LINERs but with optical periodic variability at twice the binary frequency due to rejected streams shock heating the cavity wall. Compiling a catalog of LINERs and searching for optical variability with LSST may be a promising method to detect binary activity in LLAGN. 

While this work does not account for magnetic fields, recent magnetohydrodynamic (MHD) simulations suggest that toroidally magnetized CBDs can produce magnetically arrested (MAD) accretion flows, in which large-scale poloidal flux accumulates around each black hole (\citealt{Most:2025}). The Blandford-Znajek process (\citealt{Blandford:1977}) can then power dual relativistic jets from the cavity, producing radio and X-ray emission. Hence, non-accreting SMBHBs may appear as radio-loud, X-ray-bright but UV/optical-faint AGN with LINER spectra as a result of low UV ionizing flux. 

\subsection{Asymmetric Double-Peaked Line Emission}
\begin{figure}
    \centering
    \includegraphics[width=\linewidth]{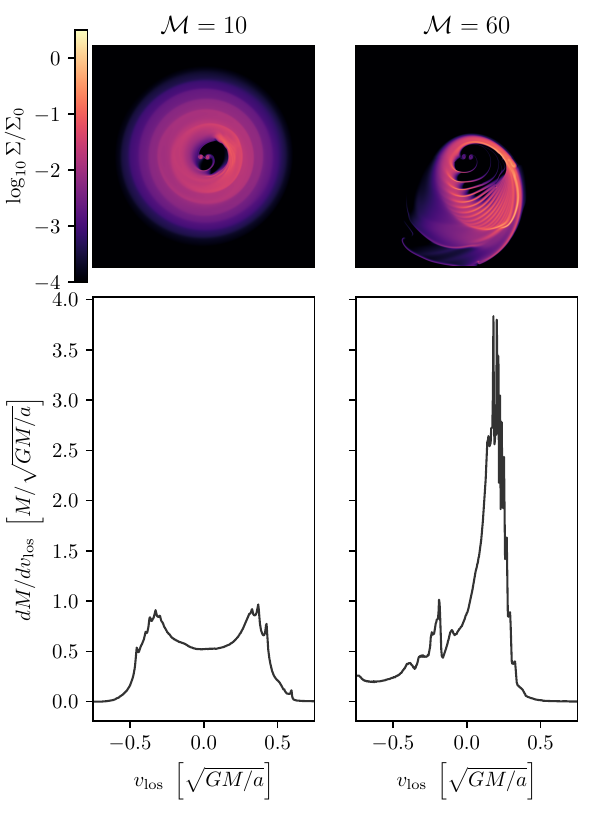}
    
    \caption{(Top) Density snapshots of ring $R_0=2a$ and $\mathcal{M}=\{10,\ 60\}$ at similar phases of precession. (Bottom) Corresponding velocity line profiles along the line of sight (LOS) chosen along the positive $x$-direction.}
    \label{fig:line_profiles}
\end{figure}
When observed near edge-on, accretion disks emit a double-peaked line profile as a result of Doppler shift from gas approaching and receding at opposite sides of the disk (\citealt{Eracleous:1994}; \citealt{Eracleous:1995}). While circular disks exhibit approximately symmetric peaks, eccentric disks can produce enhanced asymmetry in relative strength as well as frequency separation of the peaks (\citealt{Eracleous:1995}). Disk eccentricity has been proposed as a possible explanation for asymmetric double-peaks in AGN line profiles (\citealt{Strateva_2003}) and from disks evolved from TDEs (\citealt{Wevers:2022}). In the latter, the extreme eccentricity of the disk ($e=0.91\pm 0.01$) allows for measurement of the time-evolution of the double-peaks during the course of the disk's precession on $\sim$week timescales.

We have shown that CBD eccentricity growth is suppressed by infinite disks that supply angular momentum to the inner circumbinary gas. Figure \ref{fig:e_profiles} shows that these extended disks saturate at an orbital eccentricity of $e\simeq 0.05$, even for higher $\mathcal{M}$. CBRs saturate at a significantly higher orbital eccentricity; gas in ring $R_0=1a$ peaks just above $e\simeq 0.3$, and the peak saturation eccentricity rises for higher $\mathcal{M}$. Figure \ref{fig:line_profiles} illustrates the higher disk asymmetry for higher $\mathcal{M}$. The density plots (top panels) show the difference in large-scale disk eccentricity from rings of different $\mathcal{M}$. For the $\mathcal{M}=10$ ring, the significant pressure support in the gas has a circularizing effect as discussed in Section \ref{Gas_Eccentricity}. At $\mathcal{M}=60$ this effect is weak relative to stream impacts. As a result the gas settles into a disk with significant eccentricity, creating a large asymmetry in velocity at orbital apsides. The distribution $dM/dv_{\rm los}$, where $v_{\rm los}$ is the velocity along the chosen line of sight, illustrates how the eccentricity of the disk would manifest in an observed line profile. While the double-peaked line profile of $\mathcal{M}=10$ resembles that of a circular accretion disk, the asymmetric $\mathcal{M}=60$ profile is a potential signature of the eccentric disk.

In the elliptical disk model outlined by \cite{Eracleous:1995}, eccentricity affects both the separation and relative strength of double-peaked Balmer line emission profiles of AGN (see Fig. 3 in \citealt{Eracleous:1995}). \cite{Strateva_2003} show that eccentricity as low as $e\simeq 0.2$ could explain observed asymmetry in a sample of AGNs from the Sloan Digital Sky Survey, with some candidates suggesting an even more extreme $e\simeq 0.8$. If disk eccentricity in AGN is indeed driven by binaries, our findings indicate that the initial circumbinary ring must be of very small extent, with a radius comparable to the binary's semimajor axis. Our measured disk eccentricities are high enough to explain observed asymmetry in accretion disk line emission (\citealt{Strateva_2003}), if this asymmetry is indeed the result of disk eccentricity as opposed to other features such as warps, spiral shocks, or temperature variability.
\newpage
\section{Summary}\label{Summary}
Stellar and black hole binaries are shaped by their coupled interaction with gaseous environments, dictating their dynamical evolution and electromagnetic detectability. Previous works have largely studied relatively `thick' circumbinary disks (CBDs) with aspect ratio $h/r\sim \mathcal{M}^{-1} = 0.1$ and `infinite' extent. Using high-resolution hydrodynamics simulations, we studied the viscous evolution of a finite circumbinary ring (CBR) as it spreads to form a CBD around a circular, equal-mass binary. Future work is needed to study CBRs around around eccentric ($e_b \neq 0$) and unequal-mass ($q_b \neq 1$) binaries. Our suite of simulations spans ring radii in the range $R_0=1a-4a$, capturing the limit in which the infinite disk approximation is valid; and Mach numbers $\mathcal{M}=10-60$, to probe the dynamical properties of truly `thin' disks. We summarize our findings below.

We have found that accretion dynamics are insensitive to large-scale disk structure. Both finite rings and infinite disks exhibit suppressed accretion onto the binary for higher $\mathcal{M}$. While the degree of accretion suppression is sensitive to viscosity prescription, our results suggest that high-$\mathcal{M}$ systems may produce reduced accretion-driven electromagnetic luminosity. We comment on the possibility that inefficiently-accreting SMBHBs may be responsible for weak accretion flows inferred from observations of LINER spectra in low-luminosity AGN. We show that gas surrounding IMBHBs may still show variability as QPEs sourced by gas streams rejected by the binary shocking the disk inner edge. 

CBRs exhibit novel accretion periodicity, with accretion rate time series resembling quasi-triangular waves as opposed to the fiducial sawtooth pattern. CBRs surrounding circular, equal-mass binaries show lump-mediated accretion at a frequency $\sim0.1\Omega_b$. This frequency is largely insensitive to $\mathcal{M}$. If CBRs represent a common configuration for circumbinary gas distribution, our results indicate that the associated accretion periodicity (for circular, equal-mass binaries) may be less noisy when recovered from quasar light curves.

The time evolution of the gas eccentricity varies strongly with initial conditions. Infinite disks at all $\mathcal{M}$ remain nearly circular outside the central cavity, with eccentricity exponentially suppressed at large radii. In contrast, compact ($R\rightarrow1a$), colder ($\mathcal{M}\rightarrow60$) rings exhibit rapid eccentricity growth, reaching $e\simeq0.7$ and approaching a radially flat eccentricity profile. We find that the early-time growth is consistent with perturbative forcing by rejected streams, and that the prevalence of such streams in colder CBRs leads to enhanced eccentricity growth. Our results suggest that the eccentric structures we measure could account for observed asymmetries in AGN disks, if the central engine hosts a binary initially surrounded by a compact ring.

\begin{acknowledgments}
The authors thank the anonymous reviewer for a constructive report that greatly improved the quality of this work. The authors acknowledge valuable discussions with Zoltán Haiman, Magdalena Siwek and Itai Linial. This work is supported by the National Aeronautics and Space Administration (NASA) under Grant No. 80NSSC22K0822 issued through the Astrophysics Theory Program of the Science Mission Directorate.
Resources supporting this work were provided by the NASA High-End Computing (HEC) Program through the NASA Advanced Supercomputing (NAS) Division at Ames Research Center. This work was supported in part through the NYU IT High Performance Computing resources, services, and staff expertise.
\end{acknowledgments}

%% To help institutions obtain information on the effectiveness of their 
%% telescopes the AAS Journals has created a group of keywords for telescope 
%% facilities.
%
%% Following the acknowledgments section, use the following syntax and the
%% \facility{} or \facilities{} macros to list the keywords of facilities used 
%% in the research for the paper.  Each keyword is check against the master 
%% list during copy editing.  Individual instruments can be provided in 
%% parentheses, after the keyword, but they are not verified.

%% Similar to \facility{}, there is the optional \software command to allow 
%% authors a place to specify which programs were used during the creation of 
%% the manuscript. Authors should list each code and include either a
%% citation or url to the code inside ()s when available.

\software{JAX \citep{jax2018github}, astropy, matplotlib, pandas, SciPy.}

%% Appendix material should be preceded with a single \appendix command.
%% There should be a \section command for each appendix. Mark appendix
%% subsections with the same markup you use in the main body of the paper.

%% Each Appendix (indicated with \section) will be lettered A, B, C, etc.
%% The equation counter will reset when it encounters the \appendix
%% command and will number appendix equations (A1), (A2), etc. The
%% Figure and Table counter will not reset.

\appendix
\section{Estimating Black Hole Binary Populations with LISA Event Rates}\label{QPE_calc}
Our goal is to use future LISA event rates to constrain the number of intermediate-mass ($M\sim 10^4 M_\odot$) black hole binaries (IMBHBs) with periods consistent with observed QPEs ($P\sim \mathrm{hours-weeks}$). The following estimate is valid as long as the inspiral of the IMBHB is GW-dominated, i.e. if the characteristic inspiral inspiral time $\tau$ is less than a Salpeter time $t_{\rm sal} = 4.5 \times 10^7$ years. The period of the binary orbit is $P=2\pi \sqrt{a^3 / (GM)}$. Equivalently, we can write $a = (GM)^{1/3}(2\pi)^{-2/3} P^{2/3}$. Assuming a circular, equal-mass binary, the inspiral time is given by Eq (5.9) from \cite{Peters:1964},
\begin{align*}
    \tau &= a_0^4 \frac{5}{64} \frac{c^5}{G^3 M^3} \\
    &= \left( \frac{5}{64} \right) (2\pi)^{-8/3} c^5 (GM)^{-5/3} P^{8/3},
\end{align*}
which can be rearranged to solve for the initial period $P$ given an inspiral time $\tau$,
\begin{align}
    P &= 2\pi \tau^{3/8} \left(\frac{5}{64}\right)^{-3/8} c^{-15/8} (GM)^{5/8} \notag \\
    &= 13.3\ \textrm{days}\ \left( \frac{M}{10^4 M_\odot} \right)^{5/3} \left( \frac{\tau}{4\times10^7\ \textrm{yr}} \right)^{3/8}.
\end{align}

Let $N(P, t)$ be the number of IMBHBs with periods less than $P$. Assuming a stationary population, $\dot{N}=0$. The rate of mergers is 
\begin{align*}
    \mathcal{R}_0 \equiv \dot{P} \frac{dN}{dP},
\end{align*}
where $\mathcal{R}_0$ can be probed experimentally by LISA. Here $\mathcal{R}_0$ is also the ``flux'' of $n\equiv \frac{dN}{dP}$ where
\begin{align*}
    \frac{\partial{n}}{\partial t} + \frac{\partial f}{\partial P} = 0,
\end{align*}
is an advection equation for $n$; $f(P) \equiv \dot{P} n(p)$. In steady-state $f(P) = \mathcal{R}_0$ and $n(P) = \mathcal{R}_0 / \dot{P}$.\\

Taking the time derivative of $P$,
\begin{align*}
    \dot{P} = 6\pi^2 P^{-1} \frac{a^2}{GM} \dot{a}.
\end{align*}

For an equal-mass, circular binary, \cite{Peters:1964} Eq. (5.9) gives
\begin{align*}
    \dot{a} = -\frac{16}{5} \frac{G^3 M^3}{c^5 a^3}
\end{align*}

Then, we have
\begin{align*}
    \dot{P} &= -\left( \frac{96}{5} \right) 2^{2/3} \pi^{8/3} \frac{(GM)^{5/3}}{c^5} P^{-5/3}.
\end{align*}

The binary population is
\begin{align}
    N(P) &= \int_0^P n(P') dP' \notag \\
    &\approx 4\times 10^{7} \left( \frac{\mathcal{R}_0}{1\ \mathrm{yr^{-1}}} \right) \left( \frac{M}{10^4 M_\odot}\right)^{-5/3} \left( \frac{P}{13.3\ \mathrm{days}} \right)^{8/3}.
\end{align}

%% For this sample we use BibTeX plus aasjournals.bst to generate the
%% the bibliography. The sample631.bib file was populated from ADS. To
%% get the citations to show in the compiled file do the following:
%%
%% pdflatex sample631.tex
%% bibtext sample631
%% pdflatex sample631.tex
%% pdflatex sample631.tex

\bibliography{main}{}
\bibliographystyle{aasjournal}

%% This command is needed to show the entire author+affiliation list when
%% the collaboration and author truncation commands are used.  It has to
%% go at the end of the manuscript.
%\allauthors

%% Include this line if you are using the \added, \replaced, \deleted
%% commands to see a summary list of all changes at the end of the article.
%\listofchanges

\end{document}